\documentclass[12pt]{iopart}

\usepackage{graphics,iopams}

\newcommand\colorset{C}
\newcommand\colorsubset{B}
\newcommand\msg{m}
\newcommand\msgr{u}
\newcommand\pdfmsg{P}

\begin{document}

\title{Palette-colouring: a belief-propagation approach}

\author{Alessandro Pelizzola$^{1,2,3}$, Marco Pretti$^4$
\footnote[5]{To whom correspondence should be addressed
(marco.pretti@polito.it)}, and Jort van Mourik$^5$}

\address{$^1$ Dipartimento di Fisica, CNISM and Center for Computational Studies,
Politecnico di Torino -- Corso Duca degli Abruzzi 24, I-10129 Torino, Italy}

\address{$^2$ INFN, Sezione di Torino, Torino, Italy}

\address{$^3$ HuGeF Torino -- Via Nizza 52, I-10126 Torino, Italy}

\address{$^4$ CNR -- Consiglio Nazionale delle Ricerche, Istituto dei Sistemi Complessi,
and CNISM, Dipartimento di Fisica, Politecnico di Torino -- Corso Duca degli
Abruzzi 24, I-10129 Torino, Italy}

\address{$^5$ Non-Linearity and Complexity Research Group,
Aston University, Birmingham B4 7ET, UK}

\begin{abstract}
  We consider a variation of the prototype combinatorial-optimisation problem
  known as graph-colouring. Our optimisation goal is to colour the vertices of a graph
  with a fixed number of colours, in a way to maximise the number of different
  colours present in the set of nearest neighbours of each given vertex.
  This problem, which we pictorially call {\em palette-colouring},
  has been recently addressed as a basic example of problem
  arising in the context of distributed data storage.
  Even though it has not been proved to be NP complete, random search
  algorithms find the problem hard to solve.
  Heuristics based on a naive {\em belief propagation} algorithm are
  observed to work quite well in certain conditions.
  In this paper, we build upon the mentioned result, working out
  the correct belief propagation algorithm, which needs to take into
  account the many-body nature of the constraints present in this
  problem. This method improves the naive belief propagation approach,
  at the cost of increased computational effort.
  We also investigate the emergence of a satisfiable to unsatisfiable ``phase
  transition'' as a function of the vertex mean degree,
  for different ensembles of sparse random graphs in the large size
  (``thermodynamic'') limit.
\end{abstract}

\maketitle

\section{Introduction}

Graph colouring is a prototype of combinatorial optimisation or constraint
satisfaction problems~\cite{GareyJohnson1979}. It is NP-complete, so that it
can be taken as a benchmark for optimisation algorithms. Moreover, it is at the
core of a large number of technologically relevant combinatorial problems, such
as scheduling. The goal is to assign a colour to each vertex of a given graph
(with a fixed number of available colours), in such a way that no pair of
vertices connected by an edge have the same colour. Alternatively, one may be
satisfied with a suboptimal solution, i.e., minimising the number of vertex
pairs with the same colour.

A nice variant of the above problem has been recently proposed and investigated
by Bounkong and coworkers~\cite{BounkongVanmourikSaad2006,WongSaad2007}. The
variation consists in requiring that the set of colours assigned to each given
vertex and its neighbours includes all available colours. The latter problem,
which we pictorially call {\em palette-colouring}, has been suggested as a
basic example of constraint satisfaction problem arising in the context of
distributed data storage~\cite{BounkongVanmourikSaad2006}. The basic idea is as
follows. On a computer network with limited storage resources at each node, it
may be convenient to divide a file into a number of segments (colours), which
are then distributed over different nodes. Each given node should be able to
retrieve the different segments by accessing only its own and nearest neighbour
storage devices, whence the above described constraints. Even in this case, one
might be satisfied with a suboptimal solution, i.e., maximising the number of
colours present in each node neighbourhood. We note that palette-colouring has
not been proved to be NP-complete, but there are numerical evidences that it
becomes intractable for large system size~\cite{BounkongVanmourikSaad2006}.
With respect to ordinary colouring, the most relevant difference is that the
modified problem becomes easier to solve for graphs with higher, rather than
lower, vertex degrees.

In the last few years, different types of constraint satisfaction problems have
been faced by message passing techniques, among which Belief Propagation
(BP)~\cite{Yedidia2001,YedidiaFreemanWeiss2003}. BP has been originally
conceived as a dynamic programming algorithm to perform exact statistical
inference for Markov random field models defined on graphs without loops
(trees)~\cite{Pearl1986,Pearl1988}. Subsequently, it has been demonstrated to
be relatively good even for loopy graphs. Such a successful behaviour seems to
be related to the fact that actually BP is equivalent to determine a minimum of
an approximate free energy function (Bethe free energy) for a corresponding
thermodynamic system. The Bethe approximation was indeed very well known to
physicists~\cite{Bethe1935,Burley1972}, but the connection with BP is a
relatively recent result~\cite{Yedidia2001}.

In~\cite{BounkongVanmourikSaad2006}, Bounkong, van Mourik, and Saad analyse an
algorithm based on BP, comparing its performance with a variant of
Walksat~\cite{SelmanKautzCohen1994}. In particular the BP-based algorithm makes
use of beliefs averaged over several iterations, together with a common
decimation strategy. It is observed that, while Walksat works definitely better
for small graphs (100 vertices), the opposite occurs for larger (1000 vertices)
random graphs. This result is somehow related to the nature of BP itself, since
large random graphs are known to be tree-like, in the sense that the
probability of finite length loops tends to zero as the number of vertices
becomes large. Nevertheless, the BP algorithm employed
in~\cite{BounkongVanmourikSaad2006} follows a naive scheme, in which every
message provides a contribution to the probability distribution of a single
variable, taking into account a given interaction (constraint). Due to the
many-variable nature of the constraints present in the palette-colouring
problem, this scheme no longer provides the exact solution even in the case of
trees. As suggested in the cited work, the exact solution can be determined at
the cost of propagating generalised messages providing a contribution to the
joint probability distributions of pairs of nearest neighbour variables
(instead of single variable distributions). This scheme has already been used
in different works dealing with structural and spin glass models (see for
instance \cite{BiroliMezard2002} and~\cite{PagnaniParisiRatieville2003}), in
the presence of similar ``all-neighbours'' interactions. In the current paper,
we work out the pairwise BP scheme for the palette-colouring problem. As
readers more familiar with these methods may have noticed, this scheme is {\em
not} a generalised BP~\cite{YedidiaFreemanWeiss2000}. Indeed, in the
literature, the latter term usually denotes a class of algorithms computing the
minima of more refined free energy approximations (Kikuchi~\cite{Kikuchi1951},
rather than Bethe, free energies)~\cite{Pelizzola2005}. Here, however, we
derive an algorithm computing the correct Bethe approximation, which is, the
exact solution for loopless graphs. We then compare the performance of the new
BP algorithm (which we shall simply call BP from now on) to the naive one,
showing that further improvements can be obtained.

Let us note that the correct Bethe approximation has already been considered
for this problem by Wong and Saad~\cite{WongSaad2007}, in order to investigate
the emergence and nature of the satisfiable to unsatisfiable transition,
observed upon decreasing the mean vertex degree of different sparse random
graphs. In the replica-symmetry assumption, the authors of the cited paper
study average macroscopic properties of a given random graph ensemble, making
use of a numerical method of the population dynamics type. In the current work,
we mainly focus on the algorithmic properties of the message passing procedure,
and related decimation strategies. In particular, we discuss both analytical
and numerical strategies for limiting the increase of computational cost
arising from the pairwise messages. Also, in the last part of the paper, we
develop the distributional version of the message-passing scheme, which in the
literature is usually denoted as cavity method~\cite{MezardParisi2001} and used
to study random (glass-like) systems~\cite{MezardParisiVirasoro1987}. We limit
this analysis to the replica-symmetry assumption and to the simple
colour-symmetric (paramagnetic) solution. Within these simplifying hypotheses,
we compute the quenched entropy for given random graph ensembles, and estimate
the corresponding satisfiability threshold, partially recovering a result
of~\cite{WongSaad2007}.

\section{Statement of the problem and Belief Propagation}
\label{sec:statement_of_the_problem_and_BP}

We consider an undirected simple graph, whose vertices are denoted by
$i=1,\dots,N$. Our goal is to assign to each vertex $i$ a colour $x_i$ from a
given colour set $\colorset \equiv \{ 1, 2, \dots, q \}$, in a way to minimise
the cost (energy) function
\begin{equation}
  E(x_1,\dots,x_N) =
  \sum_{i=1}^N
  \eta(x_i,x_{\partial i})
  ,
  \label{eq:total_energy}
\end{equation}
where $\partial i$ denotes the {\em neighbourhood} of $i$ (i.e., the set of
vertices directly connected to $i$ by an edge), and $x_{\partial i} \equiv
\{x_j\}_{j \in \partial i}$ the array of colour variables in $\partial i$. The
elementary energy term $\eta(x_i,x_{\partial i})$ counts the number of missing
colours in the neighbourhood of $i$, including $i$ itself. A suitable
expression for the $\eta$ function is therefore
\begin{equation}
  \eta(x_1,\dots,x_n)
  = \sum_{x \in \colorset}
  \prod_{i=1}^n [1-\delta(x_i,x)] ,
  \label{eq:cluster_energy}
\end{equation}
where $\delta(x,y)$ is a Kronecker delta, and $n$ is the number of entries of
the $\eta$ function (not a-priori fixed). With the above definitions, the cost
function value is $E(x_1,\dots,x_N)=0$ if and only if the colour assignments
$x_1,\dots,x_N$ satisfy all constraints.

In the current work, we deal with this problem by studying an equivalent
``thermodynamic'' system, whose potential energy is defined by the cost
function $E(x_1,\dots,x_N)$. For energy minimisation, we consider the zero
temperature limit. The BP approach allows us to determine approximate marginals
of the equilibrium (Boltzmann) probability distribution for the colour
variables. As mentioned in the Introduction, our approximation becomes exact
when the graph is a tree. From the treatment described in
\ref{app:bp_equations}, it turns out that we can write two different marginals,
namely, the joint distribution of two colour variables on a graph edge
$p_{i,j}(x_i,x_j)$, and the joint distribution of a given colour variable
together with its neighbours $p_{i,\partial i}(x_i,x_{\partial i})$ (``cluster''
distribution), as a function of pairwise messages $m_{j \to i}(x_j,x_i)$. Each
given term $m_{j \to i}(x_j,x_i)$ may be viewed as a message sent from the
cluster $\{j,\partial j\}$ to the edge $\{i,j\}$, representing the influence of
the constraint associated to the vertex $j$ onto the colour variables of the
edge $\{i,j\}$ (some details about this interpretation are elucidated in
\ref{app:factor_graph_formalism}). In formulae, we have
\begin{eqnarray}
  p_{i,j}(x_i,x_j)
  & =
  \rme^{f_{ij}}
  \, m_{i \to j}(x_i,x_j)
  \, m_{j \to i}(x_j,x_i)
  , \label{eq:plnk} \\
  p_{i,\partial i}(x_i,x_{\partial i})
  & =
  \rme^{f_i-\beta\eta(x_i,x_{\partial i})}
  \prod_{j \in \partial i}
  m_{j \to i}(x_j,x_i)
  , \label{eq:pnod}
\end{eqnarray}
where $\beta$ is the inverse temperature, and $f_{ij}$ and $f_i$,
usually called free energy shifts (see \ref{app:bp_equations}),
can be determined by normalisation as
\begin{eqnarray}
  \rme^{-f_{ij}}
  & = \sum_{x_i,x_j} m_{i \to j}(x_i,x_j) \, m_{j \to i}(x_j,x_i)
  , \label{eq:flnk} \\
  \rme^{-f_i}
  & = \sum_{x_i,x_{\partial i}}
  \rme^{-\beta \eta(x_i,x_{\partial i})}
  \prod_{j \in \partial i} m_{j \to i}(x_j,x_i)
  . \label{eq:fnod}
\end{eqnarray}
The messages have to satisfy a set of self-consistency equations, which
basically account for compatibility between ``overlapping'' distributions. For
instance, the $\{i,j\}$ edge distribution must be a marginal of the cluster
distributions associated to both vertices $i$ and $j$. Considering the former
case, we can write
\begin{equation}
  p_{i,j}(x_i,x_j)
  = \sum_{x_{\partial i \setminus j}} p_{i,\partial i}(x_i,x_{\partial i})
  , \label{eq:compatibility}
\end{equation}
where the sum runs over the values of the array of colour variables
$x_{\partial i \setminus j} \equiv \{x_k\}_{k \in \partial i \setminus j}$,
i.e., the colour variables in the neighbourhood of $i$ except $x_j$. In fact,
we can  obtain the self-consistency equation by replacing \eref{eq:plnk}
and~\eref{eq:pnod} into the compatibility equation~\eref{eq:compatibility},
yielding
\begin{equation}
  m_{i \to j}(x_i,x_j)
  \propto
  \sum_{x_{\partial i \setminus j}}
  \rme^{- \beta \eta(x_i,x_{\partial i})}
  \prod_{k \in \partial i \setminus j}
  m_{k \to i}(x_k,x_i)
  ,
  \label{eq:rec}
\end{equation}
where a normalisation factor has been replaced by the proportionality symbol.
In order to satisfy all the necessary compatibilities, one equation of the
above form must hold for each directed edge $i \to j$. The BP algorithm solves
the set of self-consistency equations iteratively, starting from suitable
(usually random or uniform) initial conditions for the messages, until the
distance between messages at subsequent updates goes below a given threshold.
From a heuristic point of view, each message update according to~\eref{eq:rec}
is usually interpreted as a propagation process, so that in the following we
shall also denote \eref{eq:rec} as the {\em propagation equation}. For
completeness, in \ref{app:factor_graph_formalism} we also report the
propagation equations of the naive BP algorithm, which are numerically simpler.

We note that, by employing the explicit expression~\eref{eq:cluster_energy} of
the elementary energy term (cluster energy), we can significantly reduce the
computational cost of the propagation equation~\eref{eq:rec} as well. Indeed,
it turns out that the latter can be rewritten as
\begin{equation}
  m_{i \to j}(x_i,x_j)
  \propto
  \sum_{\colorsubset \subseteq \colorset \setminus x_i \setminus x_j}
  (-1+\rme^{-\beta})^{|\colorsubset|}
  \prod_{k \in \partial i \setminus j}
  \sum_{x_k \in \colorset \setminus \colorsubset}
  m_{k \to i}(x_k,x_i)
  ,
  \label{eq:rec-simp}
\end{equation}
where the outer sum runs over all the possible subsets $\colorsubset$ of the
colour set $\colorset$ without the colours $x_i,x_j$. The derivation can be
found in \ref{app:simplified_equations}. Now, we compare the computational cost
of the generic equations with respect to the simplified form. Assuming that $d$
is the degree of vertex $i$, the generic equation~\eref{eq:rec} requires
$(d-1){q}^{d-1}$ multiplications, which can be reduced to
$2{q}^{d-1}+\sum_{n=2}^{d-2}{q}^{d-n}$ by suitable (straightforward)
programming tricks. Taking into account that a trivial necessary condition for
an elementary constraint to be satisfiable is $d \geq q-1$, the leading term of
the computational cost turns out to be at least ${q}^{q-2}$. The simplified
equation~\eref{eq:rec-simp}, however, requires $(d-1){2}^{q-2}$
multiplications, which is clearly much more convenient for any $q>2$.

Finally (for completeness and future use), we also report
the simplified expression of the cluster free energy
shift~\eref{eq:fnod}
\begin{equation}
  \rme^{-f_i} =
  \sum_{x_i \in \colorset} \sum_{\colorsubset \subseteq \colorset \setminus x_i}
  (-1+\rme^{-\beta})^{|\colorsubset|}
  \prod_{j \in \partial i}
  \sum_{x_j \in \colorset \setminus \colorsubset}
  m_{j \to i}(x_j,x_i)
  ,
  \label{eq:fnod-simp}
\end{equation}
which can be obtained by an analogous derivation.

\section{Optimisation strategy and numerical results}

In this section, we define the optimisation strategy, and test its performance
on single instances of random graphs drawn from a suitable ensemble. Our
strategy involves a decimation procedure, which is analogous to that
of~\cite{BounkongVanmourikSaad2006}, but is carried out on the basis of
nearest-neighbour pair distributions $p_{i,j}(x_i,x_j)$, rather than
single-variable distributions. Given a graph and a number $q$ of available
colours, we first fix the colour of a randomly chosen vertex, in order to break
the colour permutation symmetry, and proceed as follows. We perform the first
BP run (starting from uniform messages) and determine the pair distributions
according to \eref{eq:plnk}. For each edge $\{i,j\}$, we fix the colour
variables $x_i,x_j$ at the values $\bar{x}_i,\bar{x}_j$ having the largest
joint probability, provided the latter is larger than a certain threshold. If
no probability satisfies such a condition, we only fix the pair of variables
with the largest joint probability over the whole graph. Then, we rerun BP
(starting from the previously computed messages) and iterate the above
procedure until all variables are fixed, or all constraints are satisfied (in
the latter case, non-fixed variables can be assigned a random colour). We
always set the threshold probability at $0.9$, as done
in~\cite{BounkongVanmourikSaad2006}. We observe that, in most cases, one of the
two variables chosen to be fixed has been already fixed at a previous stage of
the decimation procedure, so that, in most cases, we actually fix just one
variable for each given pair. Therefore, even though we are working with pair,
rather than single-variable distributions, we observe that choosing the same
threshold probability results in a similar decimation rate.

We now spend a few words on the precise meaning of ``fixing a variable'', as
introduced above, from the point of view of the message-passing procedure. In
the thermodynamic language, colouring a vertex is tantamount to imposing an
infinite energy penalty to all other possible colours. Thus, if we want to fix
a single variable $x_i$ to a given colour $\bar{x}_i$, we may add to the
corresponding cluster energy $\eta(x_i,x_{\partial i})$ a term $\gamma
[1-\delta(x_i,\bar{x}_i)]$, and then take the limit $\gamma \to \infty$. By the
propagation equation~\eref{eq:rec}, it is easy to see that such operations
imply that all the messages $m_{i \to j}(x_i,x_j)$, sent from the vertex $i$
(more precisely, from the cluster associated to the vertex $i$), must be
multiplied by a prefactor $\delta(x_i,\bar{x}_i)$, which basically preserves
only messages of the type $m_{i \to j}(\bar{x}_i,x_j)$. As a consequence, when
we fix the colours of two nearby vertices, it turns out that the latter no
longer need to exchange messages or, in other words, the messages remain fixed
at
\begin{equation}
  m_{i \to j}(x_i,x_j) = m_{j \to i}(x_j,x_i)
  = \delta(x_i,\bar{x}_i) \, \delta(x_j,\bar{x}_j)
  .
\end{equation}
Although such messages have no effect on the vertices $i$ and $j$ themselves,
due to the form of the propagation equation, they may still influence their
neighbourhoods $\partial i \setminus j$ and $\partial j \setminus i$.

Before presenting the results, we note that in~\cite{BounkongVanmourikSaad2006}
the authors observe that the naive BP hardly ever converges. This problem is
circumvented by computing probability distributions as ``time-averages'' over a
number of iterations, which turns out to provide sufficient information for
guiding the decimation procedure. In our scheme, the BP algorithm turns out to
converge more frequently, except in the vicinity of the satisfiability
threshold (especially after several vertices have been coloured). Convergence
may be improved by computing the message updates as convex linear combinations
between the old estimates (with coefficient $\alpha$) and the updates obtained
from the propagation equation (with coefficient $1-\alpha$). The adjustable
parameter $\alpha$ plays the role of a damping in the propagation dynamics, and
we refer to it as the {\em damping parameter}. Nevertheless, we generally find
that reaching convergence is not really necessary. Indeed, a very small number
$\nu$ of sequential updates\footnote{With reference to the propagation
equation~\eref{eq:rec}, by {\em sequential update} we mean that, in generating
a given ``output'' (left-hand side) message, one makes use of updated ``input''
(right-hand side) messages, if already available.} of all messages are
sufficient to provide the relevant information about pair probabilities, and
that a larger number of iterations does not significantly improve the overall
algorithm performance. This fact allows us to drastically reduce the
computational cost of the full procedure, although it does not affect the
complexity of a single iteration.

We are now in a position to perform a quantitative comparison with the naive BP
approach~\cite{BounkongVanmourikSaad2006}. As in the cited work, we consider a
number of available colours $q=4$ and random graphs with $N=1000$ vertices.
Graphs are generated in such a way to have vertices with two different degrees
$d=\lfloor c \rfloor$ and $d=\lceil c \rceil$, where $c$ is the mean degree.
The degree distribution, i.e., the probability of a vertex having degree $d$,
is therefore
\begin{equation}
  \rho_d = \cases
  {
    \lceil c \rceil - c    & if $d = \lfloor c \rfloor$ \\
    c - \lfloor c \rfloor  & if $d = \lceil c \rceil$   \\
    0 & otherwise
    , \label{eq:linear_distribution}
  }
\end{equation}
which we denote as {\em linear} distribution. We always assume $c
\geq q-1$, in order to avoid the appearance of vertices with
degree less than $q-1$, for which the local constraints are
necessarily unsatisfiable. We do not report results about
graphs with {\em cut-Poissonian} degree
distribution~\cite{BounkongVanmourikSaad2006}, which exhibit
analogous behaviour.

\begin{figure}[t!]
  \resizebox{165mm}{!}{\includegraphics*{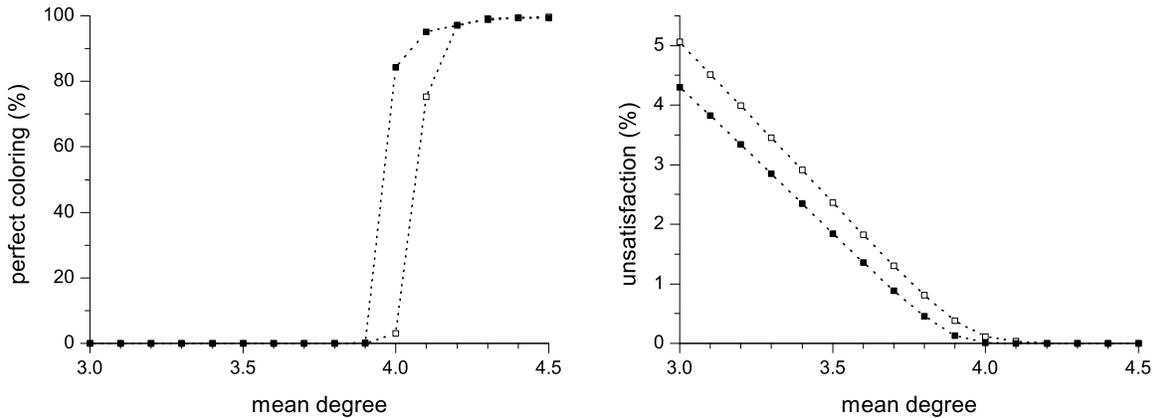}}
  \caption
  {
    Perfect colouring (left) and unsatisfaction (right) measures over 1000 graphs
    for naive BP~\cite{BounkongVanmourikSaad2006} (open squares)
    and BP with $\nu=3$ and $\alpha=0.1$ (solid squares),
    as a function of the mean degree.
    In both cases the inverse temperature used for the computation is $\beta = 10$.
  }
  \label{fig:confronto_bp}
\end{figure}
In figure~\ref{fig:confronto_bp} we report both {\em perfect colouring} and
{\em unsatisfaction} measures, over 1000 random graph samples, as a function of
the mean degree. The perfect colouring measure is simply defined  as the
fraction of samples for which the algorithm has been able to find a colour
assignment satisfying all constraints. The unsatisfaction measure counts the
fraction of missing colours per vertex, i.e. the energy per vertex divided by
the total number of colours, $E(x_1,\dots,x_N)/Nq$ ($x_1,\dots,x_N$ being the
colour assignments found by the algorithm), averaged over all samples. We can
see that the BP approach improves the naive one in both respects. The perfect
colouring measure turns out to be consistently increased in the vicinity of the
critical mean degree values, below which it rapidly vanishes. In this region,
naive BP itself was already found to work better than the Walksat-like
algorithm, analysed in~\cite{BounkongVanmourikSaad2006}.

In analogy with the ordinary colouring
problem~\cite{MuletPagnaniWeigtZecchina2002} (though with reversed role for the
mean degree~$c$), we expect that, for even lower $c$~values, our problem
becomes unsatisfiable with high probability (i.e., with probability tending to
1 in the ``thermodynamic'' $N\to\infty$ limit). We also expect the presence of
an intermediate hard-satisfiable phase in which the problem is satisfiable with
high probability but BP fails, because of a clustered structure of the solution
space (replica-symmetry
breaking)~\cite{MezardParisi2001,MezardParisiVirasoro1987,MuletPagnaniWeigtZecchina2002,KrzakalaPagnaniWeigt2004,ZdeborovaKrzakala2007}.
Accordingly, the perfect colouring probability falling down to zero is likely
to indicate the onset of such hard-satisfiable phase rather than the truly
unsatisfiable phase. We shall return to this point later. For the moment, we
observe that the BP approach definitely works better than the naive one, even
for very low $c$~values, in the (expected) unsatisfiable phase. In this region
we observe both a reduction of the unsatisfaction measure itself and of its
growth rate with decreasing~$c$.

\begin{figure}[t!]
  \resizebox{165mm}{!}{\includegraphics*{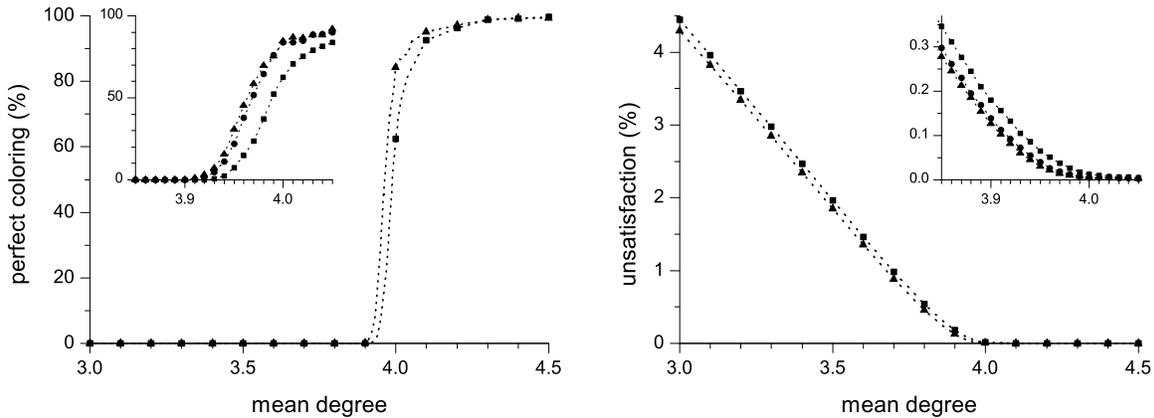}}
  \caption
  {
    Perfect colouring (left) and unsatisfaction (right) measures over 1000 graphs
    for BP with $\alpha=0.1$ and $\beta=10$, as a function of the mean degree.
    Squares, circles, triangles denote $\nu=1,2,3$, respectively.
    In the main figures, interpolation between data-points in the transition region
    has been performed by taking into account the extra data-points
    reported in the insets.
  }
  \label{fig:nitvar}
\end{figure}

Concerning the percentage of perfect colouring, we have noticed that the
performance of the algorithm is significantly affected by the number $\nu$ of
iterations per decimation step, only in a narrow region close to the critical
$c$~value.  This suggests that in this region the problem is actually more
difficult to solve. Some results about the influence of the $\nu$ parameter are
reported in figure~\ref{fig:nitvar}.
Upon increasing $\nu$, some improvement can also be observed in the
unsatisfaction measure. However, as previously mentioned, increasing $\nu$
values beyond $2$ or $3$ does not yield any further significant improvement. We
also note that a quantitatively comparable improvement of the unsatisfaction
measure is obtained by choosing a small but nonzero value of the damping
parameter $\alpha$. All the results reported in the current paper have been
obtained with $\alpha=0.1$, but it turns out that in a rather large range
($0.03\lesssim\alpha\lesssim0.3$) the average algorithm performance is
practically independent of the precise value of the damping parameter. Finally,
we note that (for $\nu\geq2$) the perfect colouring measure exhibits a slight
kink at $c=4.0$. This can be ascribed to an abrupt change in the structure of
the graph ensemble. In fact, according to the linear degree
distribution~\eref{eq:linear_distribution}, for $c=4$ all vertices have exactly
degree $4$, whereas, for $c>4$ or $c<4$ a few vertices appear with degree $5$
or $3$, respectively.

\begin{figure}[t!]
  \resizebox{165mm}{!}{\includegraphics*{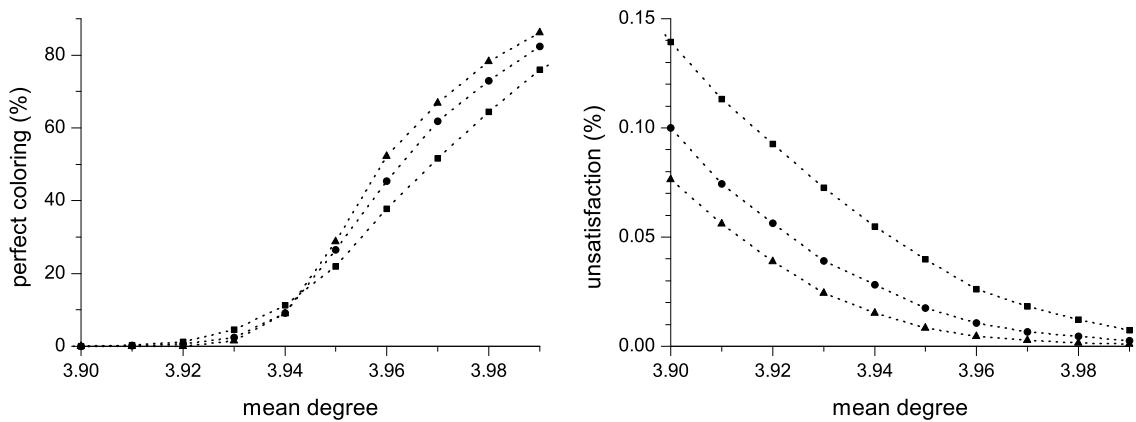}}
  \caption
  {
    Perfect colouring (left) and unsatisfaction (right) measures over 1000 graphs
    for BP with $\nu=2$, $\alpha=0.1$, and $\beta=10$, as a function of the mean degree.
    Squares, circles, triangles denote number of vertices
    $N=1000,2000,4000$, respectively.
  }
  \label{fig:nnodvar}
\end{figure}
We have also analysed the algorithm behaviour as a function of the number of
vertices~$N$. The results are reported in figure~\ref{fig:nnodvar}. We can see
that the transition in the perfect colouring probability becomes more and more
abrupt upon increasing $N$, and a cross-over point appears at a mean degree
value $c\approx3.943$. Even though the precise cross-over value may depend on
the algorithm parameters, such behaviour suggests that the transition may be
sharp (first-order-like) in the $N\to\infty$ limit. The latter conjecture is
consistent with the fact that random graphs of increasing size become more and
more tree-like, such that the BP approach is able to provide better and better
approximations. In principle, the cross-over point might be the signature of
the satisfiable to unsatisfiable transition, but, as previously mentioned, we
are rather led to identify it with the onset of the hard-satisfiable phase.
Indeed, the estimate of the satisfiability threshold, carried out in the next
section, provides further evidence in favour of the latter hypothesis.

\section{Entropy and satisfiability threshold}

In this section, we study average macroscopic properties of the BP solution
over random graph ensembles, with particular attention to the average entropy.
The latter is usually denoted as {\em quenched entropy} in statistical
mechanics language. Taking the limit $\beta \to \infty$, this quantity provides
an average measure of (the logarithm of) the number of zero energy
configurations, i.e., perfect colourings, for a given ensemble, which also
allows us to estimate the satisfiability threshold. In this context, the main
source of approximation will be the replica-symmetry assumption, since the
approximation due to BP itself is expected to be negligible in the infinite
size limit. Furthermore, we limit the analysis to BP solutions that do not
break the colour permutation symmetry (``paramagnetic'' solutions), because we
have numerical evidence that, when BP converges, no spontaneous symmetry
breaking of the solution is ever observed. Average properties of
non-paramagnetic (glass-like) solutions have been investigated
in~\cite{WongSaad2007}, but they do only appear at very low $c$ values, where
the replica-symmetry assumption is expected to break down anyway.

According to the paramagnetic ansatz, the messages are always such that
$\msg_{i \to j}(x,x)$ does not depend on $x$, and $\msg_{i \to j}(x,y)$ does
not depend on $x,y$, if $x \neq y$. This means that the only important quantity
is $\msgr_{i \to j} \equiv \msg_{i \to j}(x,x) / \msg_{i \to j}(x,y)$, i.e.,
the ratio between the ``equal colours'' message and the ``different colours''
message. Taking into account that the message normalisation is irrelevant to
all observable quantities, we can write the full message as
\begin{equation}
  m_{i \to j}(x,y)
  = 1 - (1-\msgr_{i \to j}) \, \delta(x,y)
  = \cases{\msgr_{i \to j}  & if $x=y$ \\ 1 & otherwise}
  .
  \label{eq:nobias_message}
\end{equation}
We note that in principle one could also think about the inverse ratio $\msg_{i
\to j}(x,y) / \msg_{i \to j}(x,x)$ as the relevant message, but this choice
turns out to be unfeasible, due to the nature of the constraints, favouring the
presence of different neighbouring colours. Indeed, at zero temperature, it is
easy to foresee the emergence of ``hard'' messages such that $\msg_{i \to
j}(x,x)=0$, stemming from vertices with degree $q-1$ (whose all neighbours are
forced to have a different colour), whereas we always expect $\msg_{i \to
j}(x,y) \neq 0$ for $x \neq y$ in a paramagnetic state.

Replacing \eref{eq:nobias_message} into the inner sum appearing in the
simplified propagation equation~\eref{eq:rec-simp}, we can write
\begin{equation}
  \sum_{x_k \in \colorset \setminus \colorsubset} m_{k \to i}(x_k,x_i)
  = q-|\colorsubset| -1+\msgr_{k \to i}
  ,
\end{equation}
where the term $-1+\msgr_{k \to i}$ appears because $x_i\notin\colorsubset$.
Since the sum above only depends on $\colorsubset$ via its cardinality
$|\colorsubset|$, in \eref{eq:rec-simp} we can replace the sum over
$\colorsubset$ by a sum over cardinalities, inserting suitable binomial
coefficients. Thus we finally obtain a reduced propagation equation for the
message ratios:
\begin{equation}
  \msgr_{i \to j}
  =
  \frac
  {
    \displaystyle \
    \sum_{n=0}^{q-1} {q-1 \choose n} (-1)^n
    \prod_{k \in \partial i \setminus j}
    (q-n-1+\msgr_{k \to i}) \
  }
  {
    \displaystyle \
    \sum_{n=0}^{q-2} {q-2 \choose n} (-1)^n
    \prod_{k \in \partial i \setminus j}
    (q-n-1+\msgr_{k \to i}) \
  }
  ,
  \label{eq:rec-msgr}
\end{equation}
in which we have also taken the zero temperature ($\beta \to \infty$) limit.
The cluster free energy shift can be similarly derived by replacing
\eref{eq:nobias_message} into~\eref{eq:fnod-simp}. In the zero temperature
limit, we obtain
\begin{equation}
  \rme^{-f_i} = q \sum_{n=0}^{q-1} {q-1 \choose n} (-1)^n
  \prod_{j \in \partial i} (q-n-1+\msgr_{j \to i})
  .
  \label{eq:fnod-msgr}
\end{equation}
The edge free energy shifts can be directly obtained by inserting
\eref{eq:nobias_message} into~\eref{eq:flnk}
\begin{equation}
  \rme^{-f_{ij}} = q \, (q-1+\msgr_{i \to j}\,\msgr_{j \to i})
  .
  \label{eq:flnk-msgr}
\end{equation}

We can characterise a random graph ensemble by a probability distribution of
messages $\pdfmsg(u)$. Such a distribution has to obey a functional equation
(usually known as cavity equation~\cite{MezardParisi2001}) of the following
form
\begin{equation}
  \pdfmsg(u) = \sum_d \tilde{\rho}_d
  \int \! \rmd u_1 \pdfmsg(u_1) \dots \int \! \rmd u_{d-1} \pdfmsg(u_{d-1})
  \delta(u-\hat{u}(u_1,\dots,u_{d-1}))
  ,
  \label{eq:rec-distribution}
\end{equation}
where $\hat{u}(u_1,\dots,u_{d-1})$ is the ``propagation function'' defined by
\eref{eq:rec-msgr}, and where $\tilde{\rho}_d$ is the probability of finding a
vertex of degree $d$ by choosing a random direction in a randomly selected
edge. It is easy to see that $\tilde{\rho}_d$ is related to the degree
distribution $\rho_d$ as
\begin{equation}
  \tilde{\rho}_d = \frac{d \rho_d}{c}
  .
\end{equation}
In the context of the cavity method, the replica-symmetry assumption consists
in the fact that we consider a single distribution of messages. In a
replica-symmetry breaking scenario, each propagated quantity $u_{i \to j}$
(message) would be replaced by a probability distribution defined over
different ergodic components (states)~\cite{MezardParisi2001}.

We solve the functional equation~\eref{eq:rec-distribution} numerically by a
population dynamics approach~\cite{MezardParisi2001}. In a nutshell, we
represent the distribution $\pdfmsg(u)$ by an evolving population of messages.
An elementary evolution step consists in generating a new message according to
the propagation equation~\eref{eq:rec-msgr}, making use of $d-1$ messages
randomly taken from the population, where $d$ is randomly generated according
to the $\tilde{\rho}_d$ distribution. The newly generated message replaces a
randomly selected message of the population. Due to the presence of hard
messages $u=0$ generated by degree $q-1$ vertices, we observe that the message
distribution $\pdfmsg(u)$ contains a Dirac delta peak centred in zero with
weight $\tilde{\rho}_{q-1}$.

\noindent From the message distribution, we can evaluate the average cluster
and edge free energy shifts as:
\begin{eqnarray}
  \overline{f_\mathrm{c}} & = \sum_d \rho_d
  \int \! \rmd u_1 \pdfmsg(u_1) \dots \int \! \rmd u_d \pdfmsg(u_d)
  f_\mathrm{c}(u_1,\dots,u_d)
  , \\
  \overline{f_\mathrm{e}} & =
  \int \! \rmd u_1 \pdfmsg(u_1) \int \! \rmd u_2 \pdfmsg(u_2)
  f_\mathrm{e}(u_1,u_2)
  ,
\end{eqnarray}
where the functions $f_\mathrm{c}(u_1,\dots,u_d)$ and $f_\mathrm{e}(u_1,u_2)$
are defined by \eref{eq:fnod-msgr} and~\eref{eq:flnk-msgr}. Thus we obtain the
average free energy per vertex as
\begin{equation}
  \overline{f} = \overline{f_\mathrm{c}} - \frac{c}{2} \overline{f_\mathrm{e}}
  ,
  \label{eq:mean_free_energy_per_node}
\end{equation}
where $c/2$ is the average number of edges per vertex. The above formula
directly descends from \eref{eq:total_free_energy}. Finally, since we have
incorporated a $\beta$ factor in our free energy definition, and since the
limit $\beta\to\infty$ fixes the energy at zero, the entropy per vertex is
simply $s=-\overline{f}$.

For actual calculations, we have considered random graph ensembles with the
linear degree distribution (as defined in the previous section), and with the
cut-Poissonian distribution (also considered
in~\cite{BounkongVanmourikSaad2006}), defined as
\begin{equation}
  \rho_d = \cases
  {
    \rme^{-(c-q+1)} \frac{{(c-q+1)}^{d-q+1}}{(d-q+1)!} & if $d \geq q-1$ \\
    0 & otherwise
    ,
  }
\end{equation}
where $c$ is still the mean degree. This distribution also
excludes vertices with degree smaller than $q-1$ and, hence,
trivially unsatisfiable constraints.

\begin{figure}[t!]
  \resizebox{165mm}{!}{\includegraphics*{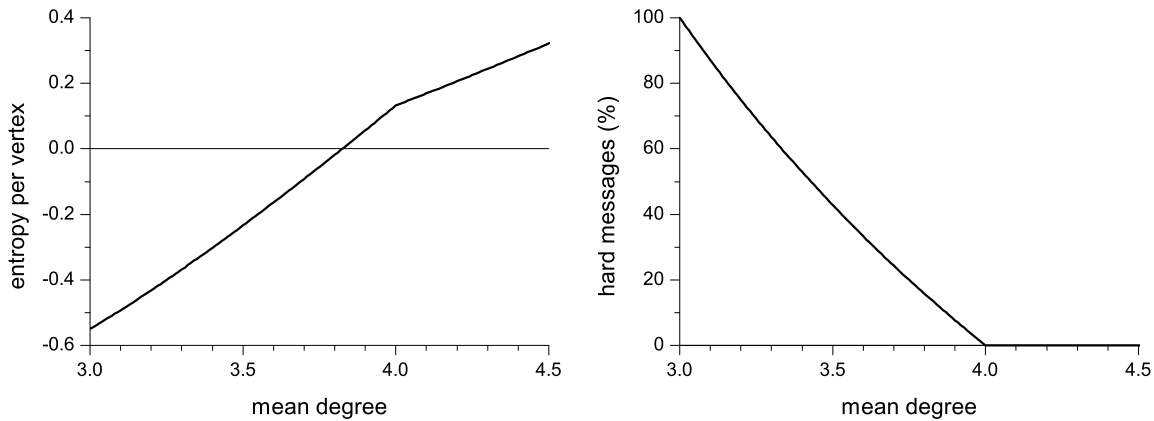}}
  \caption
  {
    Entropy per vertex (left) and fraction of hard messages (right)
    for random graphs with linear degree distribution ($q=4$),
    as a function of the mean degree $c$.
  }
  \label{fig:entropia_lin}
\end{figure}
\begin{figure}[t!]
  \resizebox{165mm}{!}{\includegraphics*{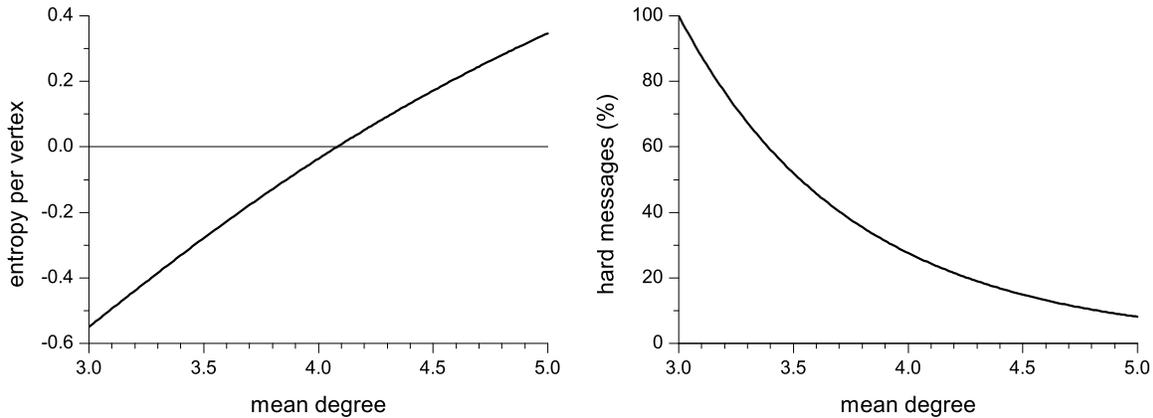}}
  \caption
  {
    The same as figure~\ref{fig:entropia_lin}
    for random graphs with cut-Poissonian degree distribution.
  }
  \label{fig:entropia_trp}
\end{figure}
In figures \ref{fig:entropia_lin} and~\ref{fig:entropia_trp} we report the
results for the two ensembles respectively. As expected, the entropy turns out
to be a monotonically increasing function of the mean degree, as the problem
becomes easier to satisfy. It is interesting to note that the fraction of hard
messages $\tilde{\rho}_{q-1}$ for the linear degree distribution turns out to
be nonzero only for $c<q$, which explains the kink observed in the entropy
function. Negative entropy identifies the unsatisfiable region (perfect
colourings are exponentially rare), whereas the zero entropy point identifies
the satisfiability threshold~$c_\mathrm{th}$. For the two ensembles, we
respectively find $c_\mathrm{th} \approx 3.825$ and $c_\mathrm{th} \approx
4.082$. As previously mentioned, we expect that these values are in fact
approximate ones, because we have neglected the possibility of replica-symmetry
breaking. Nevertheless, these values are in reasonable agreement with the
numerical estimates put forward in~\cite{BounkongVanmourikSaad2006}, namely
$c_\mathrm{th} \approx 3.8$ and $c_\mathrm{th} \approx 4.1$. As far as the
linear ensemble is concerned, we expect that our result is also analytically
equivalent to the (replica-symmetric) one by Wong and Saad~\cite{WongSaad2007},
and in fact we obtain a very good numerical agreement for the threshold value.

\section{Summary and conclusions}

In this paper, we have considered a variation of the well-known graph colouring
problem, which may be viewed as the prototype of a combinatorial optimisation
problem emerging in the context of distributed data storage. We have worked out
the BP equations for this problem, which provide the exact solution on a tree.
Due to the many-body nature of the problem, such equations turn out to be
different from the naive BP message-passing scheme, as the latter involves
messages sent to single variables, whereas the former involve messages sent to
pairs of nearest neighbour variables. Our simulations, performed on random
graphs drawn from a suitable ensemble, suggest that the new algorithm,
associated with a decimation procedure, turns out to be much more effective
than the naive BP-based algorithm. In particular, the probability of finding a
perfect colouring is significantly enhanced, especially in the vicinity of the
satisfiable-to-unsatisfiable transition. Furthermore, both the unsatisfaction
measure and its growth rate upon decreasing the average graph connectivity are
significantly reduced. This improved performance is, however, obtained at the
cost of increased computational complexity. Therefore, we have suggested two
possible ways of reducing this complexity. On the one hand, we have shown some
analytical manipulations (exploiting the particular form of the constraints)
can simplify a single iteration. On the other hand, numerical experiments have
shown that very few iterations (even a single one) provide sufficient
information to drive the decimation procedure.
We note that, in this way, our decimation procedure turns out to be somehow
``distributed'' over different BP iterations. This fact partially reminds us of
the so-called ``reinforced BP'' approach, which has been successfully exploited
to solve different combinatorial optimisation
problems~\cite{BraunsteinZecchina2006}. Although beyond the scope of the
current paper, it might be interesting to analyse the performance of the latter
method for the palette-colouring problem. Indeed, the reinforcement strategy
would replace the decimation procedure, allowing for a fully decentralised
implementation of the algorithm, which might make it actually appealing from a
practical/technical point of view.

\noindent From a more theoretical perspective, we have applied the cavity
method to investigate the satisfiable-to-unsatisfiable transition, which
appears upon decreasing the average graph connectivity. Limiting this analysis
to the replica-symmetry assumption, we have observed that the threshold
connectivity seems to be significantly displaced with respect to the value
observed in the numerical experiments. As previously mentioned, this fact
suggests that the breakdown of the algorithm may occur because of the onset of
a hard-satisfiable phase. It would also be interesting to investigate this
possibility, making use of the cavity method at the level of 1-step replica
symmetry breaking~\cite{MezardParisi2001}, along the lines of several works
dealing with the ordinary colouring
problem~\cite{MuletPagnaniWeigtZecchina2002,KrzakalaPagnaniWeigt2004,ZdeborovaKrzakala2007}.

\appendix

\section{Belief Propagation equations}
\label{app:bp_equations}

The BP equations can in general be derived from a very simple recipe. One first
``fakes'' that the graph is a tree and then formally applies the equations
obtained for such a case to a generic graph. This derivation also provides a
heuristic argument explaining why the method generally works better for graphs
with a tree-like structure.

\begin{figure}[t!]
  \resizebox{160mm}{!}
  {
    \includegraphics*{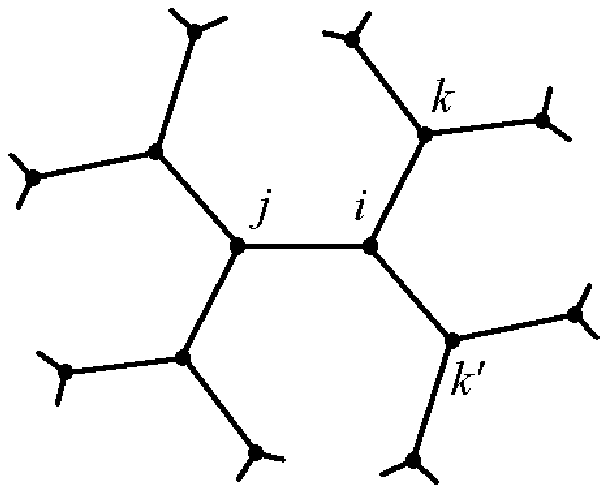}
    \includegraphics*{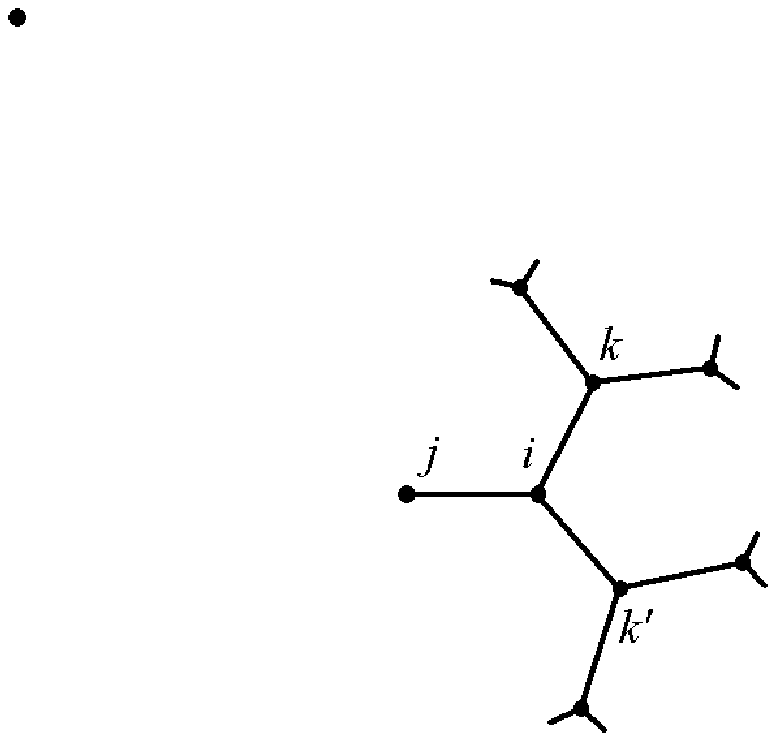}
    \includegraphics*{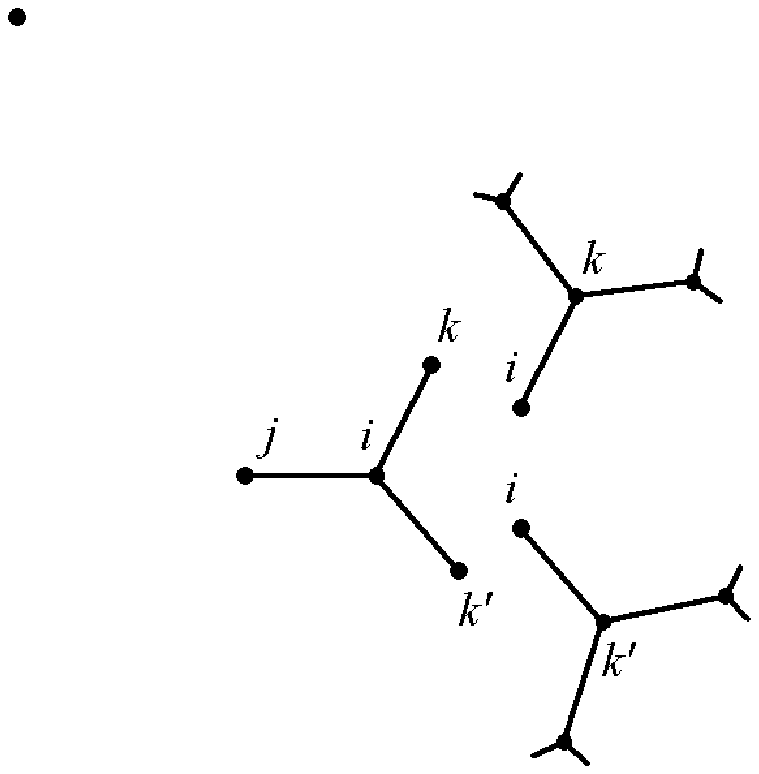}
  }
  \caption
  {
    Tree graph (left), disconnected branch $i \to j$ (centre),
    and decomposition of the latter into subbranches $k \to i$,
    for $k \in \partial i \setminus j$,
    plus the elementary cluster associated to $i$ (right).
  }
  \label{fig:trees}
\end{figure}

According to the Boltzmann law, the joint probability distribution of all the
colour variables can be written as
\begin{equation}
  p(x_1,\dots,x_N) = \rme^{F - \beta E(x_1,\dots,x_N)}
  ,
  \label{eq:boltzmann}
\end{equation}
where $E(x_1,\dots,x_N)$ is the energy function~\eref{eq:total_energy}, $\beta$
is the inverse temperature, and $F$ is the free energy (times~$\beta$), which
can be determined by normalisation. Following our ``fake assumption'', we can
consider, for each edge $i \to j$ (defined with a direction), the branch
growing from the root vertex $j$ towards $i$, disconnected from the remainder
of the system (see figure~\ref{fig:trees}).
We can thus define a partial energy function $E_{i \to j}(x_{i \to j})$,
obtained by summing the elementary interaction energies only for vertices in
the branch, except the root vertex. Since our elementary interaction energies
couple together clusters of variables including each vertex and all its
neighbours, each partial energy function depends on the array of all colour
variables in the branch including the root vertex. We denote this array by
$x_{i \to j}$. Now, each disconnected branch can be ideally studied as an
independent subsystem, whose Boltzmann probability distribution turns out to be
\begin{equation}
  p_{i \to j}(x_{i \to j})
  = \rme^{F_{i \to j} - \beta E_{i \to j}(x_{i \to j})}
  ,
  \label{eq:boltzmann_branch}
\end{equation}
where $F_{i \to j}$ denotes the corresponding free energy. Note that it is
possible to decompose the partial energy of the given branch $i \to j$ into a
sum of the partial energies of its subbranches $k \to i$, for all $k \in
\partial i \setminus j$, plus the elementary interaction energy associated to
$i$ (see figure~\ref{fig:trees}):
\begin{equation}
  E_{i \to j}(x_{i \to j}) = \eta(x_i,x_{\partial i}) +
  \sum_{k \in \partial i \setminus j} E_{k \to i}(x_{k \to i})
  .
  \label{eq:branch_energy}
\end{equation}
We also define a free energy shift $f_{i \to j}$ as the
difference between the free energy of the $i \to j$ disconnected
branch and the sum of free energies of its (disconnected)
subbranches, i.e.,
\begin{equation}
  F_{i \to j} = f_{i \to j} +
  \sum_{k \in \partial i \setminus j} F_{k \to i}
  .
  \label{eq:free_energy_shift_iter}
\end{equation}
From \eref{eq:boltzmann_branch}, \eref{eq:branch_energy}, and
\eref{eq:free_energy_shift_iter}, we can write
\begin{equation}
  p_{i \to j}(x_{i \to j}) = \rme^{f_{i \to j} - \beta \eta(x_i,x_{\partial i})}
  \prod_{k \in \partial i \setminus j} p_{k \to i}(x_{k \to i})
  ,
\end{equation}
which provides a relationship between the Boltzmann distribution of the $i \to
j$ disconnected branch and those of its (disconnected) subbranches. Defining
the messages $m_{i \to j}(x_i,x_j)$ as marginals of a corresponding branch
distribution $p_{i \to j}(x_{i \to j})$ over the variables $x_j$ and $x_i$
(respectively, the root vertex and its first neighbour in the branch) we
finally obtain the self-consistency equation~\eref{eq:rec}.

We still have to show how messages can determine cluster and edge marginals of
the full Boltzmann distribution~\eref{eq:boltzmann}. As in our previous
manipulations, we observe that, for each given vertex~$i$, it is possible to
write the total energy function~\eref{eq:total_energy} as a sum of partial
energies of the disconnected branches $j \to i$, for all $j \in \partial i$,
plus the elementary interaction energy associated to $i$:
\begin{equation}
  E(x_1,\dots,x_N) = \eta(x_i,x_{\partial i}) +
  \sum_{j \in \partial i} E_{j \to i}(x_{j \to i})
  .
  \label{eq:total energy_node}
\end{equation}
Defining also the free energy shift $f_i$ as the difference
between the total free energy $F$ and the sum of the disconnected
branch free energies, for all the possible branches growing from
vertex $i$, i.e.,
\begin{equation}
  F = f_i + \sum_{j \in \partial i} F_{j \to i}
  ,
  \label{eq:free_energy_shift_node}
\end{equation}
from \eref{eq:boltzmann}, \eref{eq:boltzmann_branch}, \eref{eq:total
energy_node}, and \eref{eq:free_energy_shift_node}, we easily obtain
\begin{equation}
  p(x_1,\dots,x_N) = \rme^{f_i - \beta \eta(x_i,x_{\partial i})}
  \prod_{j \in \partial i} p_{j \to i}(x_{j \to i})
  .
\end{equation}
Now, the cluster distribution $p_{i,\partial i}(x_i,x_{\partial i})$ for each
vertex $i$ can be derived as a suitable marginal of $p(x_1,\dots,x_N)$. By this
marginalisation, we obtain \eref{eq:pnod}. As far as edge marginals are
concerned, we have to consider a different decomposition of the total energy
function. Namely, for each edge $\{i,j\}$, the former can be written as a sum
of two contributions from respectively the branch starting from $j$ towards $i$
and the one starting from $i$ towards $j$:
\begin{equation}
  E(x_1,\dots,x_N) = E_{i \to j}(x_{i \to j}) + E_{j \to i}(x_{j \to i})
  .
  \label{eq:total energy_link}
\end{equation}
We define the free energy shift $f_{ij}$ as the
difference between the total free energy $F$ and the sum of the
free energies of the disconnected branches mentioned above, i.e.,
\begin{equation}
  F = f_{ij} + F_{i \to j} + F_{j \to i}
  .
  \label{eq:free_energy_shift_link}
\end{equation}
From \eref{eq:boltzmann}, \eref{eq:boltzmann_branch}, \eref{eq:total
energy_link}, and \eref{eq:free_energy_shift_link}, we obtain
\begin{equation}
  p(x_1,\dots,x_N) = \rme^{f_{ij}} p_{i \to j}(x_{i \to j}) p_{j \to i}(x_{j \to i})
  .
\end{equation}
Evaluating the edge distribution $p_{i,j}(x_i,x_j)$ as a marginal of
$p(x_1,\dots,x_N)$, we obtain \eref{eq:plnk}.

Finally, we determine the total free energy as a function of the free energy
shifts. First we sum both sides of \eref{eq:free_energy_shift_node} over all
vertices~$i$, and both sides of \eref{eq:free_energy_shift_link} over all
edges~$\{i,j\}$. Then we subtract the latter equation from the former. It is
easy to see that, on a tree, the number of vertices equals the number of edges
plus one, such that the left-hand side of the resulting equation turns out to
be exactly $F$. Furthermore, in the right-hand side all the branch free
energies cancel out, and we obtain
\begin{equation}
  F = \sum_{i=1}^N f_i - \sum_{\{i,j\}} f_{ij}
  ,
  \label{eq:total_free_energy}
\end{equation}
where $\sum_{\{i,j\}}$ denotes the sum over all edges.

\section{Factor graph formalism}
\label{app:factor_graph_formalism}

In this appendix, we first introduce a more general form of BP equations,
defined on a {\em factor graph}~\cite{KschnischangFreyLoeliger2001}. Then, we
show that from this form one can derive both the naive BP equations
of~\cite{BounkongVanmourikSaad2006} and the BP equations of the current paper
by two different factor graphs associated to the same problem.

\begin{figure}[t!]
  \resizebox{160mm}{!}
  {
    \includegraphics*{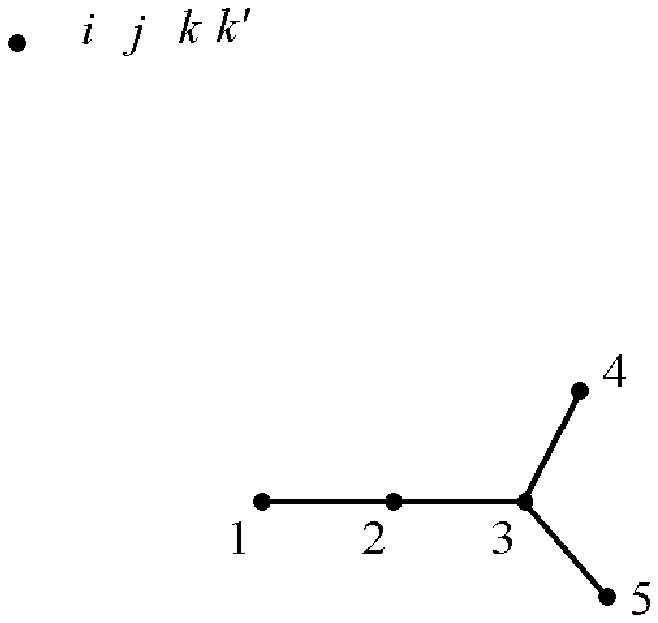}
    \includegraphics*{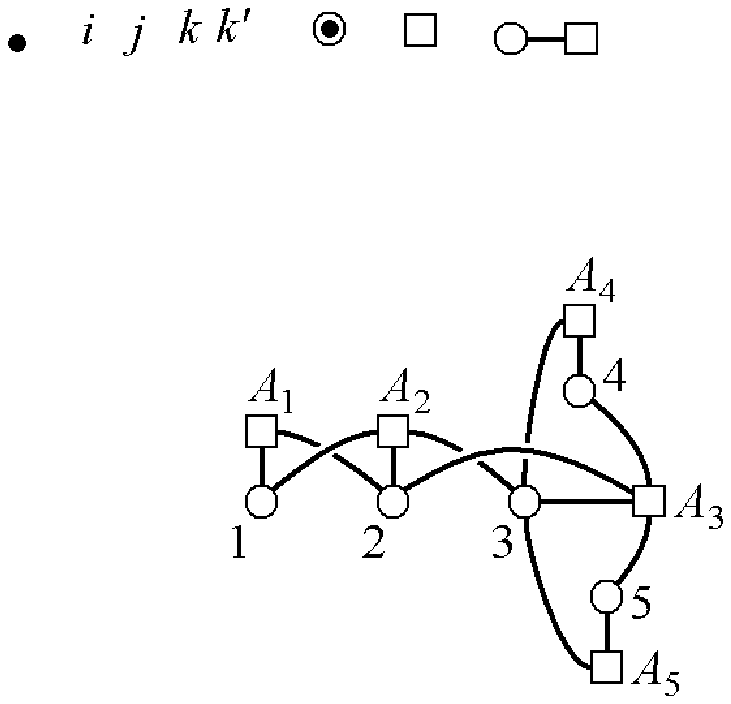}
    \includegraphics*{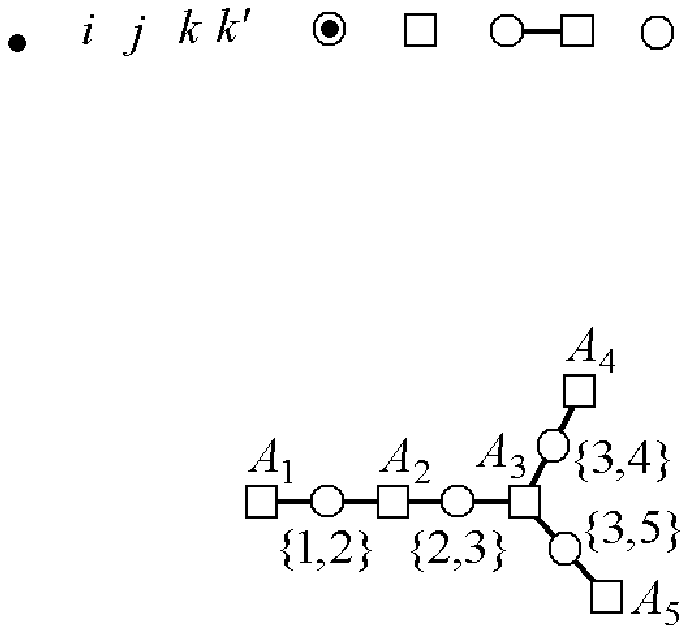}
  }
  \caption
  {
    A simple undirected graph (left), and the related factor graphs
    giving rise to naive BP (centre) and BP (right).
    Open circles and squares denote variable and function nodes, respectively.
    The labels are explained in the text.
  }
  \label{fig:graphs}
\end{figure}

A factor graph is a bipartite graph, whose left- and right-side vertices are
usually referred to as {\em variable nodes} and {\em function nodes}. The
notion of factor graph is meant to describe the structure of the energy
function, whose independent variables (i.e., the configuration variables of the
corresponding thermodynamic system) are associated to the variable nodes. A
function node connected to a number of variable nodes represents an elementary
interaction among the corresponding variables. Let $V$ denote the set of all
the variable nodes, such that each node $v \in V$ is associated with a
configuration variable $x_v$. Let also $A \subseteq V$ denote any subset
(cluster) of variable nodes, and let $x_A \equiv \{x_v\}_{v \in A}$ denote the
array of the associated configuration variables. We can thus write the energy
function as
\begin{equation}
  E(x_V) = \sum_{A \in \mathcal{F}} \epsilon_A(x_A) ,
\end{equation}
where $\epsilon_A(x_A)$ denotes the elementary interaction energy among the
variables in the cluster $A$ (cluster energy), whereas the sum runs over the
set $\mathcal{F}$ of all the interacting clusters. In what follows, the same
label $A$ denotes both a function node and the cluster of variable nodes
connected to it. An example of factor graphs describing the energy function of
a palette-colouring problem is sketched in figure~\ref{fig:graphs}.

When the factor graph is a tree, an argument similar to that in
\ref{app:bp_equations} allows one to write marginals of the Boltzmann
distribution as follows:\\
-- For each variable node $v \in V$ we have the marginal:
\begin{equation}
  p_v(x_v) =
  \rme^{f_v}
  \prod_{\stackrel{\scriptstyle A \in \mathcal{F}}{A \ni v}}
  m_{A \to v}(x_v)
  ,
  \label{eq:pvar}
\end{equation}
where the product runs over all the clusters $A$ to which $v$ belongs (i.e. all
the function nodes connected to $v$), $m_{A \to v}(x_v)$ is a {\em
function-to-variable} message, and $f_v$ is a free energy shift (ensuring
normalisation).\\
-- For each cluster $A \in \mathcal{F}$, we have the marginal:
\begin{equation}
  p_A(x_A) =
  \rme^{f_A - \beta\epsilon_A(x_A)}
  \prod_{v \in A} w_{v \to A}(x_v)
  ,
  \label{eq:pfun}
\end{equation}
where $f_A$ is a free energy shift, and where $w_{v \to A}(x_v)$ is a {\em
variable-to-function} message:
\begin{equation}
  w_{v \to A}(x_v) =
  \prod_{\stackrel{\scriptstyle A' \in \mathcal{F} \setminus A}{A' \ni v}}
  m_{A' \to v}(x_v)
  ,
  \label{eq:inverse_message}
\end{equation}
a product of the messages sent to $v$ from all connected function nodes except
$A$.\\
As shown in Section~\ref{sec:statement_of_the_problem_and_BP}, one can
derive the propagation equations by imposing compatibility between overlapping
distributions. In this case, for all $A \in \mathcal{F}$ and for all $v \in A$,
we can write
\begin{equation}
  p_v(x_v) = \sum_{x_{A \setminus v}} p_A(x_A)
  ,
  \label{eq:compatibility-factor}
\end{equation}
where the sum runs over all possible values of the variables in the cluster $A$
except $x_v$. Inserting \eref{eq:pvar},~ \eref{eq:pfun}
into~\eref{eq:compatibility-factor}, we obtain the propagation equation
\begin{equation}
  m_{A \to v}(x_v)
  \propto \sum_{x_{A \setminus v}} \rme^{-\beta\epsilon_A(x_A)}
  \prod_{v' \in A \setminus v} w_{v' \to A}(x_{v'})
  ,
\end{equation}
with the $w_{v' \to A}(x_{v'})$ defined by \eref{eq:inverse_message}. Note
that, as in \eref{eq:rec}, we have replaced the normalisation factor with a
proportionality symbol. Finally, following the argument of
\ref{app:bp_equations}, we write the total free energy as a function of the
free energy shifts as
\begin{equation}
  F = \sum_{A \in \mathcal{F}} f_A - \sum_{v \in V} (d_v-1) f_v
  ,
  \label{eq:total_free_energy_factor}
\end{equation}
where $d_v$ is the degree of the variable node $v$ in the factor graph.

\subsection*{Naive BP}

We first consider the energy function~\eref{eq:total_energy}, where the
configuration (colour) variables $x_i$ are associated with the vertices
$i=1,\dots,N$ of an ordinary graph, and the elementary interaction energy
involves a cluster made up of a vertex $i$ and all its neighbours $\partial i$.
This structure is described by a factor graph in which the variable nodes are
associated with the vertices of the original graph and the function nodes with
the clusters. We can use the same index for both the variable node $i$ and the
function node with $i$ at its centre (the cluster $A_i \equiv \{i,\partial
i\}$). Hence, each variable node $i$ receives messages $m_{A_j \to i}(x_i)$
from all the function nodes $A_j$ with $j \in \partial i$, and from $A_i$
itself. With the short-hand $m_{j \to i}$ for $m_{A_j \to i}$, omitting the
normalisation factor, \eref{eq:pvar} becomes
\begin{equation}
  p_i(x_i)
  \propto m_{i \to i}(x_i) \prod_{j \in \partial i} m_{j \to i}(x_i)
  .
  \label{eq:pvar_1}
\end{equation}
Similarly, a function node $A_i$ receives variable-to-function messages from
$i$ and all $j \in \partial i$, and the cluster distribution for $A_i$
\eref{eq:pfun} becomes
\begin{equation}
  p_{i,\partial i}(x_i,x_{\partial i})
  \propto
  \rme^{-\beta\eta(x_i,x_{\partial i})}
  w_{i \to i}(x_i) \prod_{j \in \partial i} w_{j \to i}(x_j)
  .
  \label{eq:pfun_1}
\end{equation}
We have identified $\epsilon_{A_i}(x_{A_i})$ with $\eta(x_i,x_{\partial i})$,
and $w_{j \to i}$ is short-hand for $w_{j \to A_i}$. From
\eref{eq:inverse_message}, one can see that the variable-to-function messages
take two slightly different forms, depending on whether they travel (to the
cluster $A_i$) either from the ``central'' node $i$ or from a ``peripheral''
node $j \in \partial i$. In the simplified notation, we have respectively
\begin{eqnarray}
  w_{i \to i}(x_i) & = \prod_{j \in \partial i} m_{j \to i}(x_i)
  ,  \label{eq:inverse_message_1a} \\
  w_{j \to i}(x_j) & = m_{j \to j}(x_j) \prod_{k \in \partial j \setminus i}
  m_{k \to j}(x_j)
  . \label{eq:inverse_message_1b}
\end{eqnarray}
The compatibility condition~\eref{eq:compatibility-factor}, can also be written
in two different forms. For all $i=1,\dots,N$, $j \in \partial i$, we have
respectively:
\begin{eqnarray}
  p_i(x_i) & = \sum_{x_{\partial i}} p_{i,\partial i}(x_i,x_{\partial i})
  , \\
  p_i(x_i) & = \sum_{x_j,x_{\partial j \setminus i}} p_{j,\partial j}(x_j,x_{\partial j})
  .
\end{eqnarray}
Using \eref{eq:pvar_1} and~\eref{eq:pfun_1}, this in turn gives rise to two
different propagation equations:
\begin{eqnarray}
  m_{i \to i}(x_i)
  & \propto
  \sum_{x_{\partial i}} \rme^{-\beta\eta(x_i,x_{\partial i})}
  \prod_{j \in \partial i} w_{j \to i}(x_j)
  , \\
  m_{j \to i}(x_i)
  & \propto
  \sum_{x_j,x_{\partial j \setminus i}} \rme^{-\beta\eta(x_j,x_{\partial j})}
  w_{j \to j}(x_j) \prod_{k \in \partial j \setminus i} w_{k \to j}(x_k)
  .
\end{eqnarray}
These equations, together with \eref{eq:inverse_message_1a}
and~\eref{eq:inverse_message_1b}, are identical (apart from the notation) to
the naive BP equations presented in~\cite{BounkongVanmourikSaad2006}. From
figure~\ref{fig:graphs} one sees that even when the original graph is a tree,
the corresponding factor graph contains short loops, and the naive BP equations
are not exact.

\subsection*{Current BP}

We now consider an alternative form of the energy
function~\eref{eq:total_energy} by introducing:\\
(i) a variable $x_i^j$ for each vertex-neighbour pair $(i,j\in\partial
i)$ (a kind of ``replica'' of $x_i$);\\
(ii) a constraint imposing that all replicas of $x_i$ are
equal for each vertex $i$.\\
The constraints can be realised by assigning infinite energy penalties
to configurations we want to be forbidden. Assuming $\gamma\to\infty$, we
define
\begin{equation}
  \fl
  E(\{x_1^j\}_{j \in \partial 1},\dots,\{x_N^j\}_{j \in \partial N}) =
  \sum_{i=1}^N \left[ \eta(x_i^*,\{x_j^i\}_{j \in \partial i})
  + \gamma\,\chi(\{x_i^j\}_{j \in \partial i}) \right]
  ,
  \label{eq:total_energy_modified}
\end{equation}
where the function $\chi(\cdot)$ returns $1$ when its entries are not all
equal, and $0$ otherwise, whereas $x_i^*$ means that the replica index is
irrelevant. Note that the allowed (finite energy) configurations can be
directly mapped onto the configurations of the original system, and also have
the same Boltzmann weights. This does not depend on the specific form of the
cluster energy function $\eta(\cdot)$, but only on the fact that each vertex of
the original graph interacts (at most) with all its neighbours. With these
definitions, each edge $\{i,j\}$ of the original graph can be naturally
associated with the pair of variables $\{x_i^j,x_j^i\}$ (the $j$-replica of
$x_i$ and the $i$-replica of $x_j$). Moreover, the structure of the modified
energy function~\eref{eq:total_energy_modified} is described by a factor graph
in which the variable nodes $v$ correspond to the edges $\{i,j\}$ of the
original graph, while the function nodes $A$ now correspond to the clusters of
interacting edges $A_i \equiv \{\{i,j\}|j \in \partial i\}$.
Figure~\ref{fig:graphs} shows that, when the original graph is a tree, this
factor graph is also one, and every variable node $\{i,j\}$ has degree $2$, so
that it only receives messages from the function nodes $A_i$ and $A_j$. Using
$m_{i \to j}$ as short-hand for $m_{A_i \to \{i,j\}}$, \eref{eq:pvar} becomes
\begin{equation}
  p_{\{i,j\}}(x_i^j,x_j^i)
  = \rme^{f_{\{i,j\}}} \,
  m_{i \to j}(x_i^j,x_j^i) \, m_{j \to i}(x_j^i,x_i^j)
  .
  \label{eq:pvar_2}
\end{equation}
The variable-to-function messages \eref{eq:inverse_message} are simply
\begin{equation}
  w_{i \to j}(x_i^j,x_j^i) = m_{i \to j}(x_i^j,x_j^i)
  ,
\end{equation}
where $w_{i \to j}$ is short-hand for $w_{\{i,j\} \to A_j}$. Finally,
the cluster distribution~\eref{eq:pfun} is
\begin{equation}
  p_{A_i}(\{x_i^j,x_j^i\}_{j \in \partial i})
  = \rme^{f_{A_i} - \beta \eta(x_i^*,\{x_j^i\}_{j \in \partial i})
  - \beta \gamma \chi(\{x_i^j\}_{j \in \partial i})}
  \prod_{j \in \partial i} m_{j \to i}(x_j^i,x_i^j)
  ,
  \label{eq:pfun_2}
\end{equation}
where the cluster energy $\epsilon_A(x_A)$ has been replaced with the
elementary term of \eref{eq:total_energy_modified}. Discarding forbidden
configurations (dropping replica indices), \eref{eq:pvar_2} is equivalent
to~\eref{eq:plnk}, and, since all the $\chi$-terms vanish, \eref{eq:pfun_2} is
equivalent to~\eref{eq:pnod}. This is sufficient to derive the propagation
equation~\eref{eq:rec}, as shown in
Section~\ref{sec:statement_of_the_problem_and_BP}. Finally, the free
energy~\eref{eq:total_free_energy_factor} is equivalent
to~\eref{eq:total_free_energy}, as all variable nodes of the factor graph have
degree $2$.

\section{Simplified equations}
\label{app:simplified_equations}

In this appendix, we derive the simplified forms \eref{eq:rec-simp}
and~\eref{eq:fnod-simp} of the propagation equation~\eref{eq:rec} and the free
energy shift~\eref{eq:fnod}, respectively. Both derivations are based on
similar manipulations. We consider the elementary energy
term~\eref{eq:cluster_energy} associated to vertex~$i$, and note that it can be
written in an alternative form for each given choice of a neighbour vertex $j
\in \partial i$:
\begin{equation}
  \eta(x_i,x_{\partial i})
  = \sum_{x \in \colorset \setminus x_i \setminus x_j}
  \prod_{k \in \partial i \setminus j} [1-\delta(x_k,x)]
  ,
\end{equation}
where the sum runs over the colour set~$\colorset$, excluding the colours $x_i$
and~$x_j$ (if $x_i=x_j$, just one colour is excluded). Since the product in the
equation above can only take the values $0$ and~$1$, we can write the
corresponding Boltzmann factor as
\begin{equation}
  \rme^{-\beta\eta(x_i,x_{\partial i})}
  =
  \prod_{x \in \colorset \setminus x_i \setminus x_j}
  \biggr\{ 1 - (1-\rme^{-\beta})
  \prod_{k \in \partial i \setminus j} [1-\delta(x_k,x)]
  \biggr\}
  ,
\end{equation}
and expand the outer product
\begin{equation}
  \rme^{-\beta\eta(x_i,x_{\partial i})}
  =
  \sum_{\colorsubset \subseteq \colorset \setminus x_i \setminus x_j}
  (-1+\rme^{-\beta})^{|\colorsubset|}
  \prod_{x \in \colorsubset}
  \prod_{k \in \partial i \setminus j}
  [1-\delta(x_k,x)]
  ,
\end{equation}
where the sum runs over all the possible subsets $\colorsubset$ of the colour
set $\colorset \setminus x_i \setminus x_j$. Then, we exchange the two
products, expand the product over~$x$ (taking into account that every product
of two or more deltas vanishes), and use the fact that
\begin{equation}
  \sum_{x \in \colorset}\delta(x_k,x)=1
  .
\end{equation}
We finally obtain
\begin{equation}
  \rme^{-\beta\eta(x_i,x_{\partial i})}
  =
  \sum_{\colorsubset \subseteq \colorset \setminus x_i \setminus x_j}
  (-1+\rme^{-\beta})^{|\colorsubset|}
  \prod_{k \in \partial i \setminus j}
  \sum_{x \in \colorset \setminus \colorsubset}
  \delta(x_k,x)
  .
  \label{eq:boltzmann_final}
\end{equation}
The propagation equation~\eref{eq:rec} for a given vertex~$i$ generates an
outgoing message $m_{i \to j}(x_i,x_j)$ as a function of the set of incoming
messages $m_{k \to i}(x_k,x_i)$ (where $k \in \partial i \setminus j$).
Replacing the final expression for the Boltzmann
factor~\eref{eq:boltzmann_final} into this equation, we readily obtain the
simplified propagation equation~\eref{eq:rec-simp}.

As far as the free energy shift~\eref{eq:fnod} is concerned, we rewrite the
elementary energy term~\eref{eq:cluster_energy} in yet another form, namely,
\begin{equation}
  \eta(x_i,x_{\partial i})
  = \sum_{x \in \colorset \setminus x_i}
  \prod_{j \in \partial i} [1-\delta(x_j,x)]
  .
\end{equation}
In this case the sum runs over the colour set~$\colorset$, excluding only the
colour $x_i$. A totally analogous derivation allows us to write
\begin{equation}
  \rme^{-\beta\eta(x_i,x_{\partial i})}
  =
  \sum_{\colorsubset \subseteq \colorset \setminus x_i}
  (-1+\rme^{-\beta})^{|\colorsubset|}
  \prod_{j \in \partial i}
  \sum_{x \in \colorset \setminus \colorsubset}
  \delta(x_j,x)
  ,
\end{equation}
which, plugged into~\eref{eq:fnod}, yields~\eref{eq:fnod-simp}.

\end{document}